\documentclass[structabstract]{aa}  
\usepackage{graphicx}
\usepackage{txfonts}
\usepackage{booktabs}
\usepackage{footnote}
\usepackage{xspace}
\usepackage{url}

\def\teff{$T_\mathrm{eff}$}                 
\def\ms{\hbox{\,m\,s$^{-1}$}}               
\def\m2s2{\hbox{\,m$^{2}$\,s$^{-2}$}}       
\def\kms{\hbox{\,km\,s$^{-1}$}}             
\def\gcm3{\hbox{\,g\,cm$^{-3}$}}            
\def\vsini{\hbox{$v$\,sin\,$i_{\star}$}}    
\def\Msun{\hbox{$M_{\odot}$}}               
\def\Rsun{\hbox{$R_{\odot}$}}               
\def\Mjup{\hbox{$\mathrm{M}_\mathrm{Jup}$}} 
\def\Rjup{\hbox{$\mathrm{R}_\mathrm{Jup}$}} 
\def\Rearth{\hbox{$\mathrm{R}_\mathrm{Earth}$}} 
\def\kepler{\emph{Kepler}}                  

\begin{document}
   \title{Confirmation of an exoplanet using the transit color signature: 
         Kepler-418b, a blended giant planet in a multiplanet system}

   \author{B.~Tingley\inst{1}
          \and
          H.~Parviainen\inst{2,3}
          \and
          D.~Gandolfi\inst{4,5}
          \and
           H.~J.~Deeg\inst{2,3}
          \and
          E.~Palle\inst{2,3}
          \and
          P.~Monta\~n\'es Rodriguez\inst{2,3}
          \and
          F.~Murgas\inst{2,3}
          \and
          R. Alonso\inst{2,3}
          \and
          H.~Bruntt\inst{1}
          \and
          M.~Fridlund\inst{6}
          }

   \institute{Stellar Astrophysics Center,
	Institut for Fysik og Astronomi,
              Aarhus Universitet,
              Ny Munkegade 120,
              8000 Aarhus C, Denmark 
              \email{tingley@phys.au.dk}
         \and
              Instituto de Astrof\'{i}sica de Canarias,
              C/ V\'{i}a L\'{a}ctea, s/n,
              38205 - La Laguna (Tenerife), Spain
         \and
              Dpto. de Astrof\'isica,
              Universidad de La Laguna,
              38206 - La Laguna (Tenerife), Spain
         \and
              INAF - Osservatorio Astrofisico di Catania, 
              Via S. Sofia, 78, 95123 Catania, Italy
	\and
	     Landessternwarte K\"onigstuhl, Zentrum f\"ur Astronomie der 
	Universit\"at Heidelberg, K\"onigstuhl 12, D-69117 Heidelberg, Germany 
         \and
              Institute for Planetary Research, DLR,
	    Rutherfordstrasse 2
             12489 Berlin, Germany
             }

   \date{Received October 15, 2011; accepted XXXXX, 2011}

\abstract
   {}
  {We announce confirmation of Kepler-418b, one
  of two proposed planets in this system. This is the first confirmation
  of an exoplanet based primarily on the transit color signature technique.}
   {We used the \kepler\ public data archive combined with multicolor
   photometry from the Gran Telescopio de Canarias and radial
   velocity follow-up using FIES at the Nordic Optical Telescope
   for confirmation\footnote{GTC $g'$ and $z'$ photometry and NOT-FIES
   radial velocity measurements are only available in electronic form
   at the CDS via anonymous ftp to cdsarc.u-strasbg.fr (130.79.128.5)
  or via http://cdsweb.u-strasbg.fr/cgi-bin/qcat?J/A+A/}.}
   {We report a confident detection of a transit color signature that
  can only be explained by a compact occulting body, entirely ruling
  out a contaminating eclipsing binary, a hierarchical triple, or a
  grazing eclipsing binary. Those findings are corroborated by our
  radial velocity measurements, which put an upper limit of 
  $\sim$1\,\Mjup\ on the mass of Kepler-418b. We also report that the host
  star is significantly blended, confirming the $\sim10$\% light 
  contamination suspected from the crowding metric in the \kepler\ 
  light curve measured by the \kepler\ team. We report detection of an
  unresolved light source that contributes an additional $\sim40$\% to
  the target star, which would not have been detected 
  without multicolor photometric analysis. The resulting planet-star 
  radius ratio is $0.110\pm0.0025$, more than 25\%
  more than the $0.087$ measured by \kepler\, leading to a radius of
  $1.20\pm0.16$\,\Rjup\ instead of the 0.94 \,\Rjup\
  measured by the \kepler\ team.}
  {This is the first confirmation of an exoplanet candidate based primarily
  on the transit color signature, demonstrating that this technique is viable from
  ground for giant planets. It is particularly useful for planets with long periods such
  as Kepler-418b, which tend to have long transit durations. While this
  technique is limited to candidates with deep transits from the ground, it
  may be possible to confirm earth-like exoplanet candidates with a
  few hours of observing time with an instrument like the James Webb Space
  Telescope. Additionally, multicolor photometric analysis of transits 
  can reveal \emph{unknown} stellar neighbors and binary companions that do not
  affect the classification of the transiting object but can have a very significant
  effect on the perceived planetary radius.}

\titlerunning{First confirmation using a transit color signature: Kepler-418b}

   \maketitle

\section{Introduction}

The \kepler\ mission (Borucki et al. \cite{bor2010}) has made some of
the most fascinating discoveries in exoplanetary science, finding planets as small as Mercury
(e.g., Kepler-37b, Barclay et al. \cite{bar2013}), planets in the habitable zone of their host star
(e.g., Kepler-22b, Borucki et al. \cite{bor2012}), many confirmed multi-planet systems exhibiting transit 
timing variations (e.g., Kepler-11, Lissauer et al. \cite{lis2011}), and detecting and releasing to the public thousands of other 
yet-unconfirmed candidates (Batalha et al \cite{bat2013}). The remaining unconfirmed candidates
contain as much potential for interesting discoveries as the 
confirmed exoplanets, if not more so. However, most of the easy 
cases have already been resolved, meaning that the remaining candidates present an ever-increasing challenge to validate them as exoplanets. 

In the absence of transit timing variations, which allow for the confirmation of multiplanet systems
without additional follow-up observations (Holman \& Murray \cite{hol2005}), astronomers have relied
heavily on spectroscopic measurements of radial velocities to confirm transiting exoplanet
candidates. These observations reveal the mass function of the system, which, when combined with
the mass of the star, yields the mass of the transiting body -- a very straightforward way of 
eliminating non-planetary false positives. However, this is not always a simple matter. Some eclipsing binaries are contaminated by another light source in such a way that their
eclipses appear to be very similar to exoplanetary transits (known colloquially as {\it blends}), but they
do not necessarily produce an observable radial velocity signal, leading to confusion over
whether a given system is a blend or an exoplanetary system with a mass below the detection
threshold. The aim of detecting low-mass planets leads to the expenditure of additional telescope time on such
systems in an effort to reduce the detection threshold. In multiple instances, extensive follow-up
observations returned no conclusive results (e.g., CoRoT's SRc01\,E1\,2630, Erikson et al. \cite{eri2012})
Further complicating matters, the amplitude of the radial velocity (RV) signals decreases as  
period increases and planetary mass decreases and may
fall below the level of the intrinsic stellar RV noise (i.e., the so-called ``jitter'') due to 
magnetic activity in stars, typically in the range of 1--10\,$m\,s^{-1}$ (e.g., Wright \cite{wri2005}). 
Even short-period rocky planets around very bright stars can be enormously difficult to detect with RVs,
resulting in controversial detections, such as the Earth-sized planet in a 3.4-day orbit around $\alpha$\,Cen. 
(Dumusque et al. \cite{dum2012}; Hatzes \cite{hat2013}).

Given these limitations, alternative techniques for determining the nature of transiting exoplanet
candidates are being developed and tested. In this paper, we present a confirmation of 
Kepler-418b based primarily on multicolor photometry. In addition to the confirmation, we also detect a large amount of blending in this system from an unresolved source, which would not have been detected without multicolor photometry and leads to a significant 
revision of the planetary parameters. In Sect.\,2, we discuss the multicolor photometric method. In Sect.\,3,
we describe the target and the \kepler\ photometry. The ground-based photometric observations and spectral observations used in the analysis are described in Sect.\,4 and Sect.\,5, respectively. In Sect.\,6, we present the results of the analysis, the implications of which we discuss in Sect.\,7.

\section{Method}

The idea of using multicolor photometry of a transit to reject
false positives is an old one, dating back to the first practical discussion of the transit method by Rosenblatt (\cite{ros1971}).
A colorimetric signature is created during transits due to an interplay
between the size of the occulting body and differential stellar limb
darkening -- the light of the disk of a star is more centrally concentrated
in the blue than in the red. Analysis of this signature reveals the true ratio of
radii of such a system, regardless of the level of contamination,
with small bodies producing a signature that is distinct
from grazing eclipsing binaries, triple systems, and blends,
as long as the transit/eclipse is not extremely grazing, as the
interplay between the small size of the planet relative to the star manifesting as
blueward "spike" at ingress/egress.
(Tingley \cite{tin2004}, Tingley et al. \cite{tin2014}). This technique can
therefore eliminate eclipsing binaries, even if they are contaminated spatially
by unresolved stars that distort the shape of the observed light curve (CEBs).
It cannot, however, distinguish between transiting bodies that have
similar radii to giant exoplanets, such as extremely late red dwarfs and
brown dwarfs. Consequently, exoplanets that are larger than the
smallest brown dwarfs (larger than $\sim$Saturn, e.g., Burrows, Heng \& Nampaisarn
\cite{bur2011}) require a mass limit from RV for confirmation, while those that
are smaller can be confirmed using only colorimetry.
Larger candidates require additional radial velocity (RV)
observations (albeit with full knowledge that an EB/CEB is no longer
a possibility), but an upper-mass limit excluding brown dwarfs is sufficient
for confirmation.

Given the sensitivity of colorimetry to the true planet-star radius ratio of the system, this technique
cannot only identify heavily blended eclipsing binaries masquerading as
transiting exoplanets, but also improve the measured radii of \textit{bona fide} exoplanets that are blended. Light-to-moderate blending of exoplanets
is potentially a very serious problem
for studies of exoplanets as an ensemble or detailed studies of
individual planets (e.g., atmospheres) as blending cannot be identified by
monochromatic observations.

\section{Kepler-418}

We chose Kepler-418b to test this 
technique based on its long period (86.7\,days), its relatively deep (0.81\%)
and long (10.2\,hr) transit, and its faintness ($r=14.6$, $V=14.98$), all of which favor
the use of the color signature over RV for the confirmation of this system.
Also designated KOI 1089 and KIC\,3247268, it is a multi-planet candidate system,
with a second planet candidate that is smaller (depth of 0.18\%) and interior to (period of 12.2\,days)
its sibling (Borucki et al. (\cite{bor2010}). According to the \kepler\ input catalog, KIC\,3247268 is a dwarf star with an effective temperature 
of \teff=$6179\pm200$\,K, surface gravity of log\,$g=4.43\pm0.30$, metallicity of $[\mathrm{M/H}]=-0.13\pm0.50$, 
and a mass and radius of $R_\star$=$1.08\pm0.16\,M_\odot$ and  $R_\star$=$1.06\pm0.57\,R_\odot$, respectively, which leads to a planetary radius for Kepler-418b
of $9.7\pm5.2$\,\Rearth, just slightly larger than Saturn (Rowe et al. \cite{row2013}).

We based our analysis on the first 14 quarters of \kepler\
data\footnote{Available from \url{http://archive.stsci.edu/kepler}}. Of these 14 quarters, 
7 have long-cadence (LC) data and 4 have short-cadence (SC) data (Q3, Q4, Q5, Q7). 
\kepler\ did not observe Kepler-418 during 3 quarters (Q6, Q10, Q14), due to the failure of the MOD-3 module.
We estimated the point-to-point scatter of the pre-search data conditioning (PDC) 
photometry to be 2.8~mmag for the SC data and 0.6~mmag for the LC data. 

Kepler-418 is not uncontaminated. According to the Data Validation Report (DVR) for this
system\footnote{\url{http://exoplanetarchive.ipac.caltech.edu/data/KeplerData/003/003247/003247268/dv/kplr003247268-20121029204534_dvr.pdf}},
the crowding metric ranges from 0.8873 (high) to 0.9451 (low), with an average of 0.9128. 
This is apparently the result of a nearby star, namely KIC\,3247243, which lies 
15\arcsec\ West of the target and has a magnitude of r=14.3~mag. The variations 
in the crowding metric appear to be seasonal, with the highest crowding occurring during 
season 4 (Q4, Q8, Q12). This level of crowding is not negligible, so one could expect 
the transit of Kepler-418 to produce deeper transits in isolation, and therefore a larger exoplanet than 
reported by the {\it Kepler} team\footnote{\url{http://archive.stsci.edu/kepler/koi/search.php}}.
The DVR also red-flagged the photo-center variations on 
this target. These are almost certainly due to the crowding -- during transit, the 
photo-center would shift towards the primary contaminant.

The crowding can also help explain the significant even-odd transit depth
differences the \kepler\ team has attributed to this candidate during variation times
of their analysis procedure -- generally considered to be a sign of a contaminating eclipsing binary.
Kepler-418b, with a period close to 90 days, has only one transit
per \kepler\ season. Given that season 4 has a significantly higher crowding metric than the other 
seasons and season 2 is essentially missing due to the failure of the CCD, it is not surprising that
the even/odd binary test warns of a possible false positive: as all but one of the transits in the even-numbered seasons comes from the season with the highest crowding, it only stands to reason
that the even transits would have a lower apparent transit depth.

\section{GTC photometry}

\begin{figure}
 \centering
 \includegraphics[width=\columnwidth]{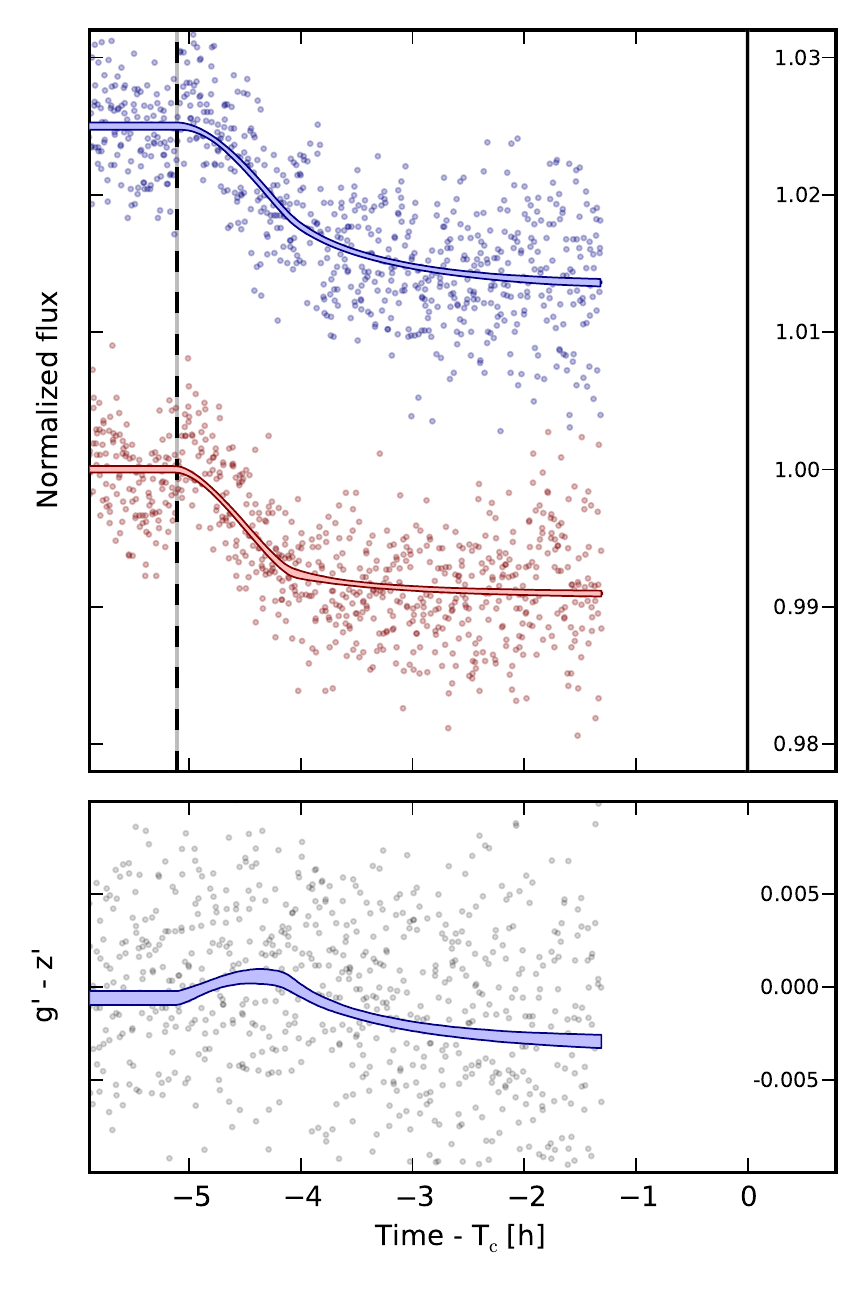}
 \caption{Kepler-418b transit observed by the GTC in Sloan g' and z' filters.
The upper figure shows the observed light curves and the 95\% 
confidence limits for the transit models, and the lower figure shows the color differences for the observed 
and modeled data in units of normalized flux, with the zero-point fitted to the off-transit data. The dashed vertical line represents the expected beginning of the transit, and the solid black line shows the fitted transit center. The 95\% confidence limits for each value are depicted as gray shaded ares, but these fall inside the line withds.}
 \label{fig:lc_gtc}
\end{figure}

Using the 10.4m Gran Telescopio Canarias (GTC), we observed a transit of Kepler-418b
on the night of 14 Aug 2011, which had first contact at UT=23:37 and a mid-transit
time of UT=04:41.  We spent about 6 hours on target, from UT=22:54 until 05:00, alternating
5 observations in $g'$ with 5 observations in $z'$, both with exposure times of 5.5 s.
These filters are well-separated chromatically ($g'=4815\pm153\AA$, $z'=969.5\pm261\AA$), so we would
expect a relatively strong transit color signature.  We obtained 760 exposures in $g'$ and
755 in $z'$, which unfortunately exhibited clear systematic distortions in the shape of the PSF. 
We extracted useful light curves from these observations using optimized aperture photometry
with the IRAF-Vaphot package (Deeg \& Doyle \cite{dee2001}), minimizing the impact
of the PSF distortions by using apertures much larger than the typical PSF, obtaining  point-to-point scatter of 5.3 and 4.8\,mmag in $g'$ 
and $z'$, respectively -- several times higher than photon noise, but adequate for our purposes.
It should be noted that other groups have reported much better performance with this instrument,
using the spectrograph
to both defocus the telescope and provide color information (Murgas et al. \cite{mur2013}).
Despite this distortion to the PSF, the resolving power of the GTC data (about 1 arcsecond) is still far superior
to \kepler\ (defocused to 10 arcseconds).
The resulting light curves can be seen in Fig.~\ref{fig:lc_gtc}. As 
expected, the transit is somewhat deeper than reported by the \kepler\ team, consistent with 
the crowding factor they reported. Moreover, a blueward spike is visible during the ingress,
visible evidence that the system has a low ratio of radii, as would be expected for a planet.

\section{NOT/FIES spectroscopy}
The candidate host star is relatively faint (r=14.6 mag), which makes obtaining
the necessary RV precision ($\sim5-10$\,\ms) to confirm an exoplanet like this
difficult. We carried out a program of RV measurements in an attempt to measure
the mass of Kepler-418b using the FIbre-fed \'Echelle Spectrograph (FIES;
Frandsen \& Lindberg \cite{frand1999}, Telting et al. \cite{telting2014}) at the 
Nordic Optical Telescope (NOT), and acquired seven high-resolution (R=67\,000) 
spectra between June and November 2012. We adopted the same instrument set-up, 
observing strategy, and data reduction described by Gandolfi et al. \cite{gandolfi2013}. 
The FIES RV measurements are listed in Table\,\ref{table:rv_observations}, along with the S/N ratio 
per pixel at 5500\,\AA. When folded to the orbital period of Kepler-418b, our measurements 
are consistent with no detectable RV variation at a level of about 40\,\ms. This puts 
an upper limit of $\sim$1\,\Mjup on the mass of the outermost transiting object, 
definitively ruling out a false positive scenario consisting of an eclipsing 
late-type red star/brown dwarf, though not any kind of blended stellar system.

Following the spectral analysis techniques outlined in Gandolfi et al. \cite{gandolfi2013}, we
used the co-added FIES spectrum, which has a S/N ratio of about 35 per pixel at
5500\,\AA, to determine the photospheric parameters of Kepler-418 spectroscopically.
The microturbulent $v_ {\mathrm{micro}}$ and macroturbulent $v_{\mathrm{macro}}$
velocities were derived using the calibration equations of Bruntt et al. (\cite{bru2010}) for
Sun-like dwarf stars. The projected rotational velocity \vsini\ was measured by
fitting the profile of several isolated metal lines. We found \teff\,$=5820\pm100$\,K,
log\,g$=4.35\pm0.10$\,dex, [M/H]~$=-0.05\pm0.10$\,dex, $v_{\mathrm{micro}}=1.1\pm0.1$\,\kms,
$v_{\mathrm{macro}}=2.6\pm0.4$\,\kms, and \vsini\,$=3.5\pm1.0$\,\kms (Table~3), slightly more
than $1\sigma$ cooler than the photospheric parameters listed in the \kepler\ input catalogue (Sect.~3) and consistent with those listed in Everett et al. (\cite{eve2013}).
We determined the mass $M_\star$, radius $R_\star$, and age of Kepler-418 by comparing the
position of the star on a log\,g-versus-\teff\ diagram with a grid of evolutionary tracks
and isochrones computed using the PAdova and TRieste Stellar Evolution Code (PARSEC; Bressan
et al. \cite{bre2012}). We derived a mass of $M_\star=0.98\pm0.08$\,\Msun\, a radius of
$R_\star=1.09\pm0.14$\,\Rsun\, and an age of $7^{+3}_{-4}$\,Gyr (Table~3), with the large
errors attributable to the error in log\,g. We use this value of the stellar radius, errors included,
to determine the radii of the planets give in Table~4.

\begin{table}[t]
  \centering 
  \caption{FIES radial velocity measurements of Kepler-418.}
  \label{rvobs}
\begin{tabular*}{\columnwidth}{@{\extracolsep{\fill}}cccc}
\toprule\toprule
\multicolumn{2}{c}{Radial velocity measurements} \\
HJD              &   RV    & $\sigma_{\mathrm RV}$ &   S/N/pixel    \\
($-$ 2\,450\,000)&  ~\kms  &   ~\kms               &    @5500\,\AA  \\
\midrule
6101.64795 & $-$18.468 &  0.042 &  11 \\
6107.65513 & $-$18.457 &  0.034 &  13 \\
6118.70532 & $-$18.426 &  0.035 &  13 \\
6122.59101 & $-$18.443 &  0.062 & ~~9 \\
6222.36569 & $-$18.472 &  0.030 &  18 \\
6242.34046 & $-$18.465 &  0.033 &  12 \\
6245.34799 & $-$18.431 &  0.030 &  13 \\
\bottomrule
\end{tabular*}
\label{table:rv_observations}
\end{table}

\begin{table*}
\centering
\caption{Kepler-418 host star parameters from spectral analysis
and from the Kepler Input Catalog.}
\begin{tabular*}{\textwidth}{@{\extracolsep{\fill}}llll}
\toprule\toprule
{\emph{\textbf{Stellar properties}}}   &  &  Spectral characterization & KIC \\
\midrule
effective temperature &[K]  & 5820$\pm$100 &  $6179\pm200$  \\
surface gravity ($\log g$) & [dex] & 4.35$\pm$0.10  &  $4.43\pm0.30$\\
metallicity  &[dex]          & $-$0.05$\pm$0.10 & $-0.13 \pm 0.50$\\
projected rotational velocity (\vsini) &[\kms] & 3.5$\pm$1.0  &    \\
microturbulent velocity$^a$ $v_ {\mathrm{micro}}$  &[\kms] & 1.1$\pm$0.1  &    \\
macroturbulent velocity$^a$ $v_ {\mathrm{macro}}$  &[\kms] & 2.6$\pm$0.4  &    \\
systemic velocity  & [\kms] & $-18.452 \pm 0.011$ & $-18.452 \pm 0.011$ \\
mass & [$M_\odot$] &  $0.98 \pm 0.08$  & $1.08 \pm 0.16$\\
radius & [$R_\odot$] &   $1.09\pm 0.14$ & $1.06 \pm 0.57$\\
age & [Gyr] & $7^{+3}_{-4}$ & \\
\bottomrule
\end{tabular*}
\label{table:stellar_parameters}
\end{table*}

\section{System characterization}
\label{sec:transit_analysis}
\subsection{Overview}
\label{sec:transit_analysis.overview}
We carried out the system characterization through simultaneous modeling of the \kepler\ 
light curves, GTC light curves, and radial velocity measurements. We used Markov Chain Monte Carlo
(MCMC)  sampling to obtain marginal posterior density estimates for the model parameters describing 
both planets and their orbits. The joint model has 31 free parameters in total, all listed in 
Table~\ref{table:parameterisation}, along with their priors. 

We base our analysis on our implementation of the Gim\'{e}nez transit shape model (Gim\'{e}nez 
\cite{Gimenez2006}), which is optimized for efficient modeling of light curves with hundreds of 
thousands data points\footnote{Available from \url{github.com/hpparvi/PyTransit}}. We super-sampled 
the long cadence model using 10 subsamples per LC~exposure to reduce the effects from the extended 
integration time, although the long transit duration of Kepler-418b reduces the impact of the low 
temporal resolution in this case (Kipping \cite{Kipping2010a}). We sampled the posterior 
distribution with \textit{emcee} (Foreman-Mackey et al. \cite{Foreman-Mackey2013}), a Python 
implementation of the Affine Invariant Markov Chain Monte Carlo sampler (Goodman \& Weare 
\cite{Goodman2010}) and initialized it with a rough global solution found with 
PyDE\footnote{Available from \url{github.com/hpparvi/PyDE}}, our implementation of the differential 
evolution global optimization algorithm~(Storn~\&~Price~\cite{sto1997}). We completed the remainder 
of the analysis using Python, SciPy, NumPy, matplotlib~(Hunter~\cite{hun2007}), and IPython.

\subsection{Blending analysis}
Blending analysis was included into the characterization by introducing a third-light source 
parameterized by the contamination factor and effective temperature of the contaminating source. The 
amount of third-light for each bandpass (Kepler, $z'$, and $g'$) was calculated for each MCMC step 
by approximating the contaminating source with a black body. A uniform prior from 0 to 0.99 was 
used on the contamination factor, while the effective temperature had a uniform prior from 2000~K 
to 15000~K. This blending parameter should not be confused with the \kepler\ crowding metric, as
we used the \kepler\ PDC data, presumably corrected for crowding, and the GTC data is uncrowded.
It therefore represents unresolved and unknown sources of contamination.

\subsection{Kepler data}
We used the \kepler\ PDC simple aperture photometry for the analysis, which corrects for the
measured crowding metric, using subsets of the light curve comprised of
38.4~h sections centered around each transit. We detrended the individual transit 
light curves using a second-order polynomial fitted to the out-of-transit points. For those
portions of the entire light curve where both LC and SC were available, we used the short
time-cadence light curves. The final SC and LC light curves contain $\sim97000$ and $\sim8000$ points, respectively.

\subsection{Parametrization and priors}
\label{sec:system_characterization.parametrization}
We chose to test two approaches to the modeling: (a) with unconstrained limb darkening 
(uninformative priors on both quadratic limb darkening coefficients) and (b) with constraining 
(informative) normal priors on the coefficients based on the theoretical values by 
Claret~et~al.~\cite{cla2013}. We constructed the normal priors such that the means match those for 
stars with $5400 < T_\mathrm{eff} < 6200$ and $4 < \log g < 5$, with standard deviations that are 
twice those of the theoretical values within the given limits. 

We use uninformative priors on all parameters for the run {\it b} to obtain the most conservative
parameter estimates possible. This was to ensure that we do not bias the blend analysis by making
assumptions about the nature of any of the objects contributing to the observed signals.

Moreover, we do not impose constraints on the eccentricity, as our RV data does not supply
information about the orbits. The transit durations of the two planets suggest that their orbits are likely eccentric,
but no sufficient constraints on the eccentricity can be set based on this. The obtained eccentricity
estimates are thus robust upper limits for the orbital eccentricities constrained by the light curves
only.

\subsection{Posterior sampling}
We initialize the \textit{emcee} sampler using a \textit{PyDE}-derived parameter vector population clumped around  the posterior maxima to reduce the burn-in time. If the posterior space is significantly multimodal, the Differential Evolution algorithm is expected to find the separate modes and allow for an MCMC analysis that spans the whole posterior space.
The posterior sampling was carried out by first running the \textit{emcee} sampler without parallel tempering iteratively with 1200 simultaneous chains 
(walkers) through a burn-in period consisting of 5 runs of 1000 steps each, after which the walkers had converged 
to sample the posterior distribution. The final sample sets for both runs consist of 3000 iterations
with a thinning factor of 100 (a value based on the chain autocorrelation length estimates), leading to 36000 independent posterior samples.

\subsection{Results}
We list the model parameter estimates for both approaches in Table~\ref{table:planet_parameters}. The values
correspond to the posterior medians and the differences between the medians and the 16th and 84th percentiles, closely 
corresponding to the posterior mean~$\pm 1\sigma$ estimates if the posteriors are close to normal. We show 95\% 
upper limits only if lower limits cannot be established (or if a lower limit equals the parameter's natural lower limit).

\begin{figure}
 \centering
 \includegraphics[width=\columnwidth]{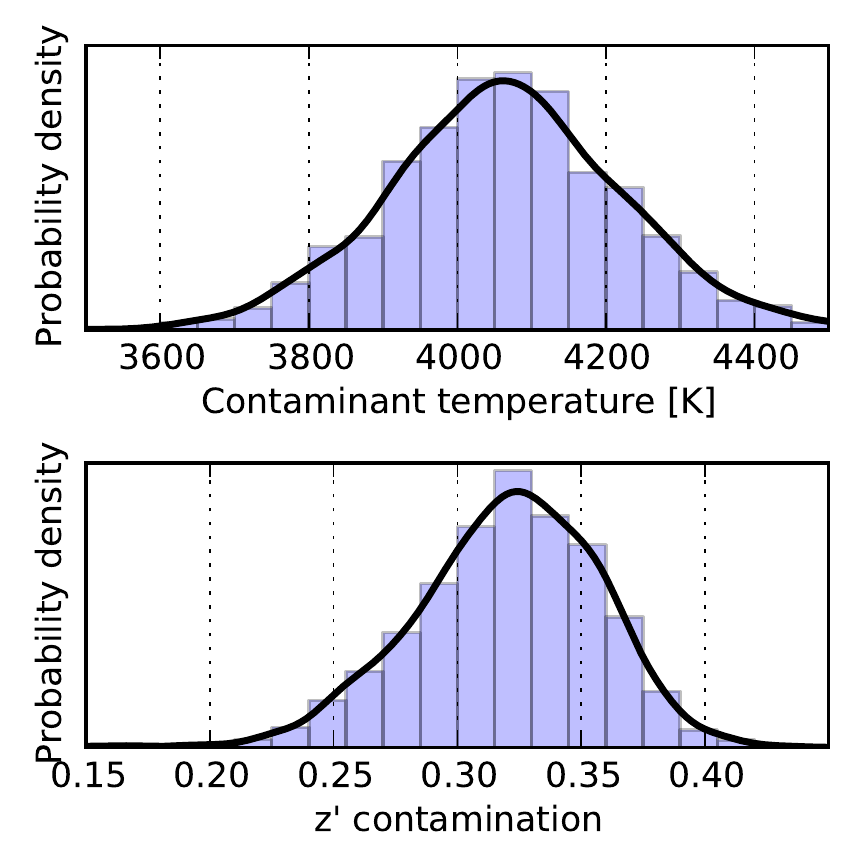}
 \caption{Marginal posterior distributions for the contaminant temperature and $z'$ contamination.}
 \label{fig:contamination_marginals}
\end{figure}

\begin{figure}
 \centering
 \includegraphics[width=\columnwidth]{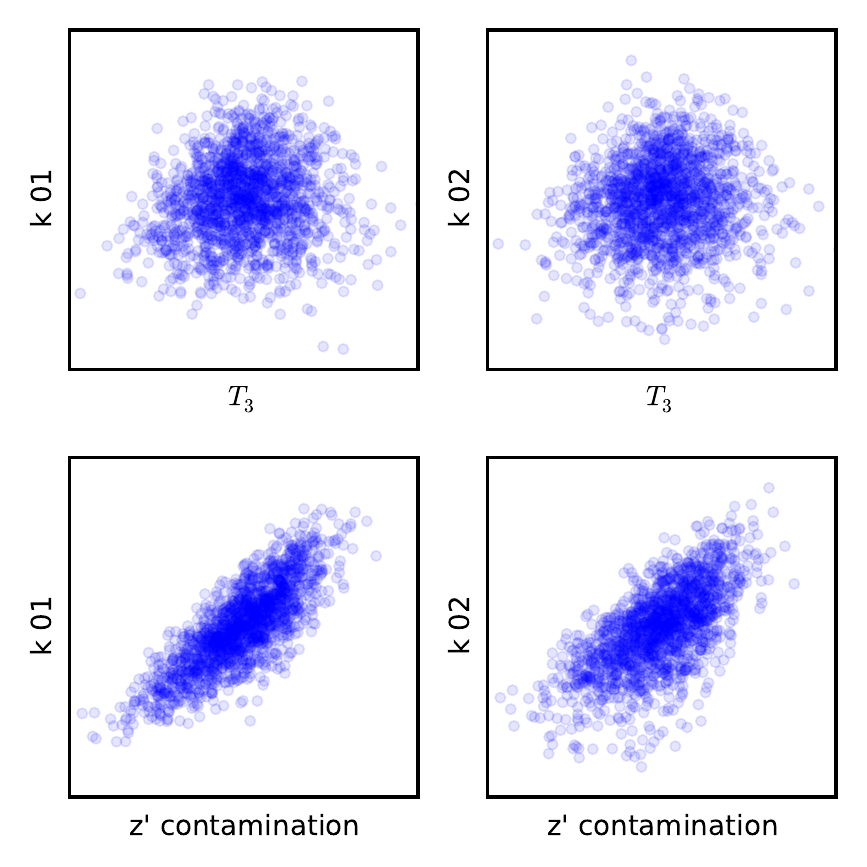}
 \caption{Joint posterior distributions for the planetary radius ratios, contamination factor, and contaminant 
 temperature.}
 \label{fig:contamination_k_dists}
\end{figure}

We also plot the marginal distributions for the contaminant temperature and $z'$ contamination in Fig.~\ref{fig:contamination_marginals}, and plot the joint posterior for the planet radius ratios, $z'$ contamination, and contaminant temperature in Fig.~\ref{fig:contamination_k_dists}. The radius ratios are 
correlated with the $z'$ contamination, and slightly correlated with limb darkening, but show no significant correlation with other fitting parameters.

The differences in the parameter estimates from the two runs are largely insignificant, well within $1\sigma$.
This is to be expected, as the estimates from the {\it a} run (unconstrained limb darkening) typically
have larger uncertainties than the estimates from the {\it b} run, and the {\it a} run
posterior contains the {\it b} run posterior. In both cases we obtain a posterior with a clear 
mode corresponding to $\sim33$\% contamination in GTC $z'$ by a 4000~K contaminating source.

\begin{table}
\begin{center}
\caption{Parameterization and priors used in the joint \kepler\ and GTC light curve and RV modelling.
The two basic prior distributions used are uniform (U) and normal (N), where the uniform distribution
is defined by its lower and upper bounds, while the normal distribution is defined by its mean and
standard deviation. The two limb darkening cases are described in detail in Sect.~\ref{sec:system_characterization.parametrization}.}
\begin{tabular*}{\columnwidth}{@{\extracolsep{\fill}}llll}
\toprule\toprule
Parameter name & & Units & Prior \\
\midrule
\multicolumn{4}{l}{\it{\textbf{System property priors for Kepler-418b}}}\\
\midrule
zero epoch & T$_{c1}$ & d & U(T$_{c1b}\pm 0.05^*$) \\
period & P$_1$ & d & U(P$_{1b}\pm 0.005^*$)\\
area ratio & k$_1^2$ & $R_\star^2$ & U($0.08^2$, $0.50^2$) \\
impact parameter & b$_1$ & & U(0.0, 0.99)\\
eccentricity & $e_1$ & & U(0.0, 0.6) \\
arg. of periastron & $\omega_1$ & & U(0, $2\pi$) \\
RV semi-amplitude & $K_1$ & km/s & U(0.0, 0.2)\\
\\
\multicolumn{4}{l}{\it{\textbf{System property priors for KOI-1089.02}}}\\
\midrule
zero epoch & T$_{c2}$ & d& U(T$_{c2b}\pm 0.05^*$) \\
period & P$_2$ & d & U(P$_{2b}\pm 0.005^*$)\\
area ratio & k$_2^2$ & $R_\star^2$ & U($0.04^2$, $0.50^2$) \\
impact parameter & b$_2$ & & U(0.0, 0.99)\\
eccentricity & $e_2$ & & U(0.0, 0.6) \\
arg. of periastron & $\omega_2$ & & U(0, $2\pi$) \\
RV semi-amplitude & $K_2$ & km/s & U(0.0, 0.2)\\
\\
stellar density & $\rho_\star$& g/cm$^3$ & U(0.1, 5.0) \\
systemic velocity & C & km/s & U(-18.4, -18.5)\\
\\
RV scatter & $e_{\rm RV}$ & km/s & U(0.01, 0.17)\\
\kepler\ LC scatter & e$_{Kl}$ & & U(0.0005, 0.0006) \\
\kepler\ SC scatter & e$_{Ks}$ & & U(0.0028, 0.0029) \\
GTC $z'$ scatter & e$_{z'}$ & & U(0.0025, 0.0045) \\
GTC $g'$ scatter & e$_{g'}$ & & U(0.0025, 0.0045) \\
GTC $z'$ baseline & z$_{z'}$ & & U(0.95, 1.05) \\
GTC $g'$ baseline & z$_{g'}$ & & U(0.95, 1.05) \\
\\
$z'$ contamination & C & & U(0.00, 0.99)\\
contaminant temperature & T$_c$ & K & U(2000, 15000)\\
\\
\midrule
\multicolumn{4}{l}{\emph{\textbf{Stellar property priors: unconstrained quadratic limb darkening}}}\\
$K$ limb darkening $u$ & $u_{a,K}$ & & U( 0.00, 1.00)\\
$K$ limb darkening $v$ & $v_{a,K}$ & & U(-0.35, 0.50) \\
$z'$ limb darkening $u$ & $u_{a,z'}$ & & U( 0.00, 1.00)\\
$z'$ limb darkening $v$ & $v_{a,z'}$ & & U(-0.35, 0.50) \\
$g'$ limb darkening $u$ & $u_{a,g'}$ & & U( 0.00, 1.00)\\
$g'$ limb darkening $v$ & $v_{a,g'}$ & & U(-0.35, 0.50) \\
\\
\midrule
\multicolumn{4}{l}{\emph{\textbf{Stellar property priors: constrained quadratic limb darkening}}}\\
$K$ limb darkening $u$ & $u_{b,K}$ & & N(0.12, 0.03)\\
$K$ limb darkening $v$ & $v_{b,K}$ & & N(0.19, 0.06) \\
$z'$ limb darkening $u$ & $u_{b,z'}$ & & N(0.32, 0.06)\\
$z'$ limb darkening $v$ & $v_{b,z'}$ & & N(0.19, 0.06) \\
$g'$ limb darkening $u$ & $u_{b,g'}$ & & N(0.64, 0.09)\\
$g'$ limb darkening $v$ & $v_{b,g'}$ & & N(0.13, 0.09) \\
\bottomrule
\\
\multicolumn{4}{l}{{\footnotesize $^*$ Transit center and orbital period values given by Borucki~\cite{bor2010}.}}
\end{tabular*}
\label{table:parameterisation}
\end{center}
\end{table}

\begin{figure*}
 \centering
 \includegraphics[width=\textwidth]{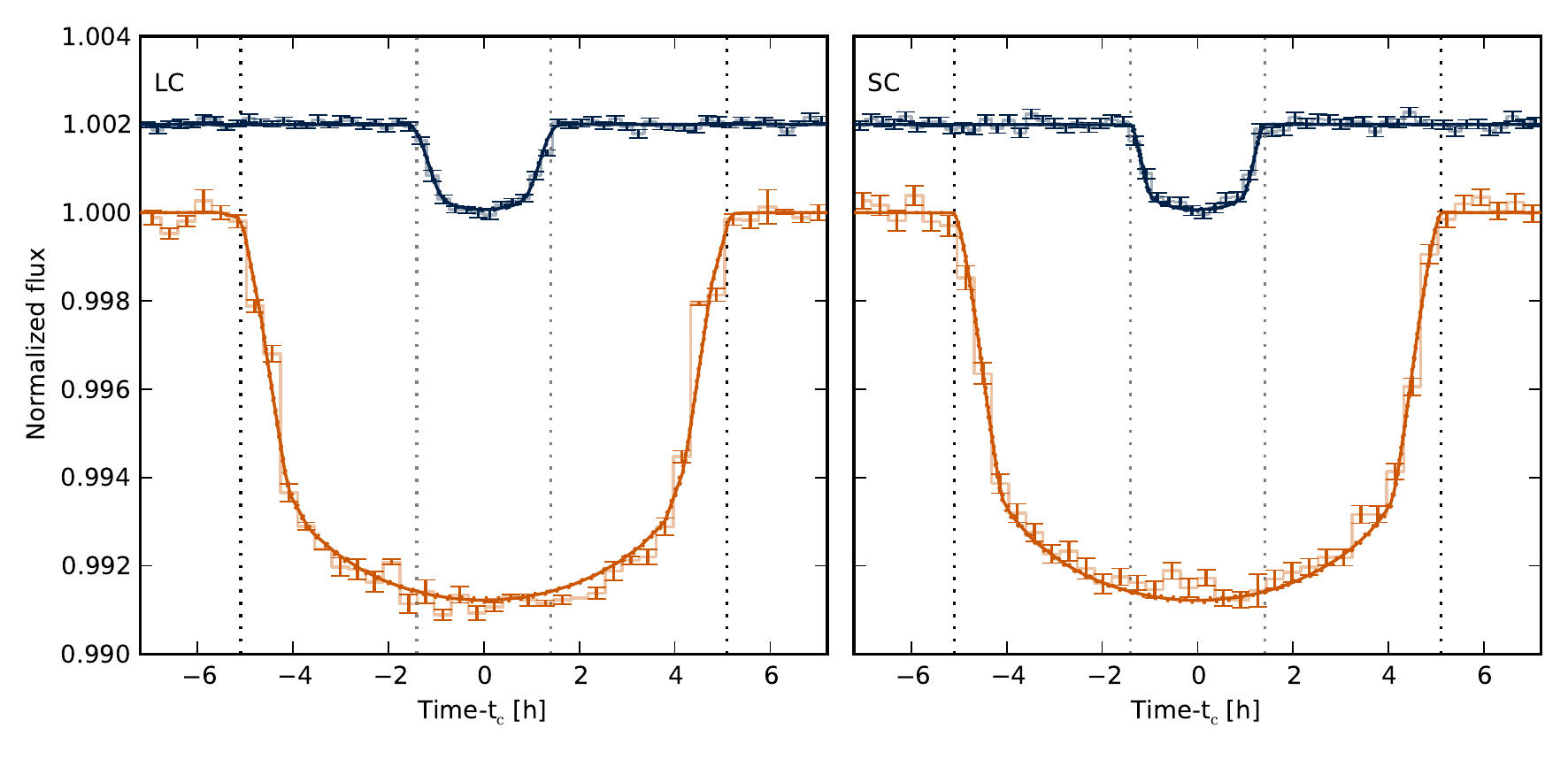}
 \caption{Phase-folded and binned  \kepler\ transit light curves showing Kepler-418b (below) and KOI-1089.02 (top).
 The solid lines represent the best-fit model and the dotted lines the first and last contact points
 for the transits. 
 The 68th percentile limits of the model predictive distributions are also shown, but fall within the line width. The left-hand panel
 shows the results fort the long-cadence data while the right-hand panel shows those for the short-cadence data.}
 \label{fig:Kepler_both_binned}
\end{figure*}

\subsection{Search for transit timing variations}
\begin{figure}
 \centering
 \includegraphics[width=0.9\columnwidth]{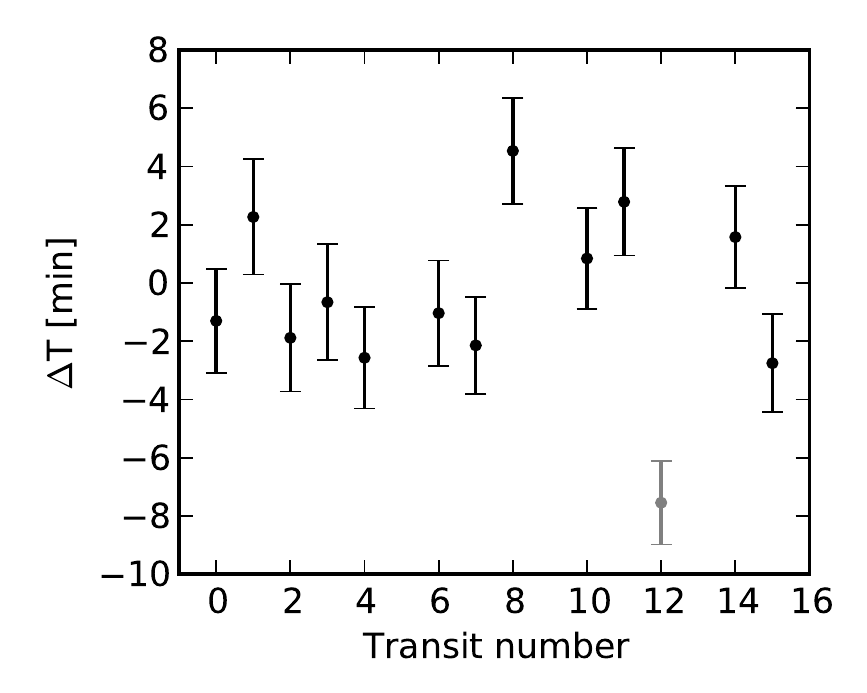}
 \includegraphics[width=0.9\columnwidth]{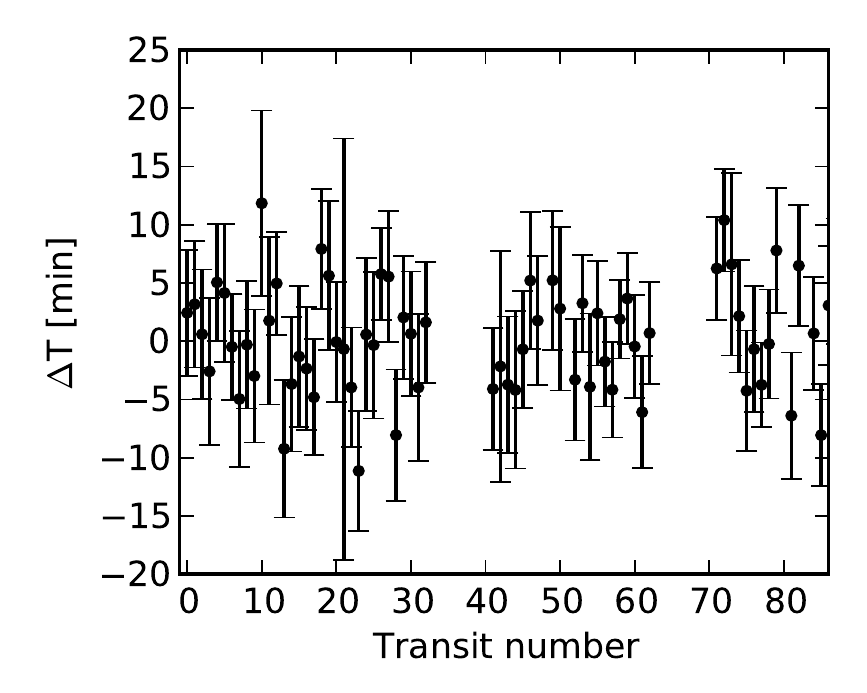}
 \caption{Transits timing variations of Kepler-418b (above) and KOI-1089.02 (below). The point drawn in gray corresponds to a transit with a dent near the $T_4$ point, and is excluded from the period fit.}
 \label{fig:ttvs}
\end{figure}

We carried out a search for variations in the transit center times (transit timing 
variations, TTVs), and show the results in Fig.~\ref{fig:ttvs}. For Kepler-418b,
the measured transit centers are completely consistent with a linear ephemeris,
once the twelfth transit is removed -- it is manifestly affected by some unknown noise source.
KOI-1089.02 also shows no sign of significant TTVs. Given the large separation between the two
planets both in period and in semi-major axis, the absence of TTVs is not unexpected. 

\begin{table*}
\centering
\caption{
    Kepler-418b planet and host star parameters from simultaneous Kepler and GTC light curve modeling. The posterior distributions for the two separate runs are described in Table~\ref{table:parameterisation}.}
\begin{tabular*}{\textwidth}{@{\extracolsep{\fill}}llll}
\toprule\toprule
\multicolumn{2}{l}{\emph{\textbf{System properties of Kepler-418b}}} & Free $u$ and $v$&  Constrained $u$ and $v$\\
Zero epoch              & [HJD - 2454833]         & $175.59750\pm 0.00066$       & $175.59734\pm 0.00068$\\
Period                  & [d]                     & $86.67856\pm 7\times10^{-5}$ & $86.67857\pm 8\times^{-5}$\\
Radius ratio            & [Rs]                    & $0.1104\pm 0.0035$           & $0.1110\pm 0.0032$\\
Impact parameter        & [-]                     & $< 0.23^a$                   & $< 0.22^a$\\
Inclination             & [deg]                   & $89.952\pm 0.041$           & $89.950\pm 0.040$  \\
Transit duration        & [h]                     & $10.224\pm 0.042$            & $10.236\pm 0.038$\\
Scaled semi-major axis  & [Rs]                    & $84.4\pm 9.5$                & $86\pm 10$\\
eccentricity            & [-]                     & $0.20\pm 0.11$               & $0.22\pm 0.12$\\
omega                   & [deg]                   & $4.35\pm 0.41$               & $4.41\pm 0.50$\\
RV semi-amplitude       & [km/s]                  & $< 0.050^a$                  & $< 0.42^a$\\
Radius                  & [R$_{Jup}$]                 & $1.20\pm 0.16$               & $1.20\pm 0.16$\\
Mass                    & [M$_{Jup}$]                 & $< 1.1^a$                    & $< 0.88^a$\\
surface gravity$^b$ ($\log g$)& [dex]             & $2.83\pm 0.47$               & $2.77\pm 0.45$\\
\\
\multicolumn{2}{l}{\emph{\textbf{System properties of KOI-1089.02}}} & Free $u$ and $v$&  Constrained $u$ and $v$\\
\midrule
Zero epoch              & [HJD - 2454833]         & $140.32089\pm 0.00090$       & $140.32099\pm 0.00087$\\
Period                  & [d]                     & $12.21826\pm 1\times10^{-5}$ & $12.21826\pm 1\times 10^{-5}$\\
Radius ratio            & [Rs]                    & $0.0577\pm 0.0019$           & $0.0577\pm 0.0018$\\
Impact parameter        & [-]                     & $0.827\pm 0.038$             & $0.827\pm 0.033$\\
Inclination             & [deg]                   & $87.60\pm 0.44$           & $88.06\pm 0.44$\\
Transit duration        & [h]                     & $2.808\pm 0.072$             & $2.827\pm 0.066$\\
Scaled semi-major axis  & [Rs]                    & $22.9\pm 2.6$                & $23.4\pm 2.9$\\
eccentricity            & [-]                     & $< 0.50^a$                   & $< 0.45^a$\\
omega                   & [deg]                   & $3.48\pm 0.55$               & $3.62\pm 0.74$\\
RV semi-amplitude       & [km/s]                  & $< 0.059^a$                  & $< 0.05^a$\\
Radius                  & [R$_{Jup}$]                 & $0.625\pm 0.083$             & $0.62557\pm 0.082$\\
Mass                    & [M$_{Jup}$]                 & $< 0.64^a$                   & $< 0.57^a$\\
surface gravity$^b$ ($\log g$)& [dex]             & $3.24\pm 0.46$               & $3.11\pm 0.46$\\
\\
\multicolumn{2}{l}{\emph{\textbf{Stellar properties from transit fit}}} & Free $u$ and $v$&  Constrained $u$ and $v$\\
\midrule
$K$ Linear ldc            & [-]        &              $0.374\pm 0.058$         &         $0.262\pm 0.025$\\
$K$ Quadratic ldc         & [-]       &               $0.14\pm 0.10$          &          $0.305\pm 0.046$\\
$z'$ Linear ldc         & [-]           &           $0.073\pm 0.088$          &        $0.246\pm 0.056$\\
$z'$ Quadratic ldc      & [-]         &             $0.36\pm 0.18$           &         $0.210\pm 0.059$\\
$g'$ Linear ldc         & [-]            &          $0.29\pm 0.16$             &       $0.606\pm 0.068$\\
$g'$ Quadratic ldc      & [-]         &             $0.79\pm 0.23$             &       $0.240\pm 0.089$\\
Stellar density         & [g/cm$^3$]     &            $1.51^{+0.68}_{-0.39}$    &        $1.62^{+0.75}_{-0.48}$\\
$K$ contamination      & [-]          &            $0.387\pm 0.037$         &         $0.393\pm 0.034$\\
$g'$ contamination      & [-]          &            $0.257\pm 0.032$         &         $0.252\pm 0.029$\\
$z'$ contamination      & [-]          &            $0.322\pm 0.037$         &         $0.335\pm 0.030$\\
Contaminant temperature & [K]       &               $4080\pm 160$         &            $4010\pm 130$\\
\\
\multicolumn{2}{l}{\emph{\textbf{Light curve noise characteristics}}} & Free $u$ and $v$&  Constrained $u$ and $v$\\
\midrule
$\sigma_K$, long-cadence$^c$  & [mmag]      &          $0.4046\pm 0.0033$    &            $0.4056\pm 0.0034$\\
$\sigma_K$, short-cadence$^c$  & [mmag]       &         $2.0230\pm 0.0044$    &            $2.0234\pm 0.0045$\\
$\sigma_{z'}$$^c$     & [mmag]           &     $3.449\pm 0.088$            &      $3.476\pm 0.083$\\
$\sigma_{g'}$$^c$      & [mmag]           &     $3.715\pm 0.095$             &     $3.715\pm 0.094$\\
\bottomrule
\\
\multicolumn{4}{l}{$^a$ {\footnotesize 95\% upper confidence limit.}}\\
\multicolumn{4}{l}{$^b$ {\footnotesize Calculated using Eq.~6.35 of Perryman~\cite{Perryman2011}.}}\\
\multicolumn{4}{l}{$^c$ {\footnotesize Point-to-point noise.}}\\
\end{tabular*}
\label{table:planet_parameters}
\end{table*}

\section{Conclusions}

We confirm that Kepler-418b is a transiting exoplanet with an upper mass
limit of 1~\Mjup. We base this confirmation primarily on the detection of
a transit color signature that is consistent with a compact occulting
body, ruling out all CEBs and high radius-ratio EBs, and secondarily
by setting an upper mass limit that rules out brown and red dwarfs, even highly
eccentric ones that would exhibit only secondary eclipses, as discussed in Santerne (\cite{san2013}). We
required only 13 hours of observations for this confirmation, 6 with the GTC
and 7 with the NOT. This is significantly less than would have been necessary
for a confirmation with RV alone with sufficient significance to rule out CEBs through
bisector analysis. For example, CoRoT-9b, a similar system, required 14 hours of
observing time with the HARPS spectrograph on the ESO 3.6m telescope for 
confirmation (Deeg et al. \cite{dee2010}). A similarly confident detection for
Kepler-418b, which is 1.3 magnitudes fainter, would require 46 hours of observations,
based on the scaling laws by Bouch, Pepe \& Queloz (\cite{bou2001}).

We detect moderate contamination from an unresolved and unknown third source
in both \kepler\ and GTC photometry. The contaminant is estimated to contribute
$\sim 33\%$ of the observed flux in $z'$.  Despite the high level of contamination,
we would not expect to be able to detect it in our spectra, due to their low signal-to-noise, and
indeed we do not.

The Kepler-418 system is an example of a blended exoplanet system. 
Multicolor photometric analysis is the only technique to date that can
definitively identify low levels of contamination. While Kepler-418b may
be one of the first for which significant contamination was detected, 
it is unlikely to be unique. Contamination has a measurable impact on
planet parameters and may even affect detailed studies of planetary
atmospheres, particularly if the contaminating star is cool and the
observations are in the infrared.

\begin{acknowledgements}
This article is based on observations made with the GTC operated on
the island of La Palma by the IAC in the Spanish Observatorio of El
Roque. The \kepler\ data presented in this paper were obtained from the
Multimission Archive at the Space Telescope Science Institute
(MAST). STScI is operated by the Association of Universities for
Research in Astronomy, Inc., under NASA contract NAS5-26555. Support
for MAST for non-HST data is provided by the NASA Office of Space
Science via grant NNX09AF08G and by other grants and contracts.
HD acknowledges funding by grant AYA2012-39346-C02-02 of the Spanish Ministry of Science and
Innovation (MICINN). Based on observations obtained with the Nordic Optical Telescope 
(NOT), operated on the island of La Palma jointly by Denmark, Finland, 
Iceland, Norway, and Sweden, in the Spanish Observatorio del Roque de 
los Muchachos of the Instituto de Astrofisica de Canarias, in time allocated 
by OPTICON by the Spanish Time Allocation Committee (CAT), through a NOT 
``fast-track'' service programme. The research leading to these 
results has received funding from the European Community's Seventh Framework 
Programme (FP7/2007-2013) under grant agreement numbers RG226604 (OPTICON) and 
267251 (AstroFit).

\end{acknowledgements}

\end{document}